\documentclass[aps,prl,reprint,superscriptaddress]{revtex4-2}
\usepackage[caption=false]{subfig}
\usepackage{graphicx}
\usepackage{amssymb}
\usepackage{xcolor}
\usepackage{soul}

\begin{document}


\title{Tunneling driven by quantum light described via field Bohmian trajectories}


\author{Sangwon Kim}
\affiliation{Department of Physics, KAIST, Daejeon 34141, Republic of Korea}

\author{Seongjin Ahn}
\email{seongjin.ahn@kaist.ac.kr}
\affiliation{Department of Physics, KAIST, Daejeon 34141, Republic of Korea}
\affiliation{Research Center for Natural Sciences, KAIST, Daejeon 34141, Republic of Korea}

\author{Denis V. Seletskiy}

\affiliation{femtoQ Lab, Department of Engineering Physics, Polytechnique Montr\'{e}al, Montr\'{e}al, QC, H3T 1J4, Canada}

\author{Andrey S. Moskalenko}
\email{moskalenko@kaist.ac.kr}
\affiliation{Department of Physics, KAIST, Daejeon 34141, Republic of Korea}


\date{\today}

\begin{abstract}
Recent realization of an intense quantum light, namely bright squeezed vacuum, opened a new perspective on quantum light-matter interaction. Several theoretical works have appeared based on coherent state expansions of quantum state of light to investigate non-classical driving of high-harmonic generation in atomic gases and solids, or free-electron dynamics, but their predictions surprisingly coincide with what one could expect from essentially classical interpretations of the light statistics. A deeper theoretical insight into the underlying physics is necessary for understanding of observed experimental findings and predicting emerging effects relying on this new configuration. Here we present a theoretical framework to describe tunneling driven by quantum light, where the properties of such light are captured by a statistical ensemble of classical fields via a hydrodynamic, also referred to as Bohmian, formulation. Generalizing the quasiclassical theory of non-adiabatic tunneling driven by classical light, a single tunneling event is described by a bundle of tunneling solutions, each driven by a classical field corresponding to one realization in the ensemble. Quantum statistics of light are thus imprinted on the measured current. Fully quantum description of light via the Bohmian trajectories of its field provides a perfect fit to the description of the electron (under-) above-barrier dynamics in terms of (complex quasiclassical) real classical trajectories, resulting in a consistent and elegant theoretical approach. To illustrate this, we consider BSV-induced electron transport from the tip to the surface in the tunneling microscope configuration demonstrating the transition from the multiphoton to the direct tunneling regime.

\end{abstract}


\maketitle


\textit{Introduction---}Interaction between light and matter has always been a subject of prospering research over the past decades, advancing fundamental physical knowledge whereas also leading to numerous technological advances. The process of high-harmonic generation (HHG) \cite{McPherson1987, L'Huillier1988, Ghimire2011, Luu2018} is arguably one of the most striking, enabling sources of extreme ultraviolet (XUV) emission \cite{Han2016}, attosecond light pulses \cite{Agostini2001, Hentschel2001}, attosecond science \cite{Krausz2009, Corkum2007, Villeneuve2018} and spectroscopy \cite{Goulielmakis2007, Eckle2008, Goulielmakis2015}.

To build on the thriving impact of the HHG phenomenon, many theoretical descriptions followed to elucidate its underlying principles. HHG is a strong-field physics phenomenon, which requires description of light-matter interaction in the non-perturbative regime. Pioneering theoretical works \cite{Kulander1992_PRA,Kulander1992} culminated in the seminal ``three-step'' model \cite{Corkum1993,Kulander1993}, which successfully described the atomic HHG process as a sequence of electron liberation via the tunnel ionization process, its classical motion under an applied optical field, and recollision with the parent ion. This theoretical framework was successful enough to explain some key characteristics of HHG, such as the frequency plateau and cutoff \cite{L'Huillier1993, Winterfeldt2008}. Further, the theory was advanced to describe the electron dynamics fully quantum-mechanically, recovering the predictions of the three-step model \cite{Lewenstein1994}. Subsequent studies paid attention to quantum properties of the involved matter part \cite{Becker1997,Salamin2006,Alon1998,Hernandez2013}, including extension to solid-state systems \cite{Vampa2014,Schubert2014, Ndabashimiye2016,Wu2016,Ghimire2019,Yue2022}, while treatment of light remained classical, constituting the so-called semi-classical light-matter interaction. This was justified since light from the utilized sources could be considered as a coherent field \cite{Glauber1963a,Glauber1963b}, with quantum corrections being negligible with respect to the instantaneous mean value of the field.

Meanwhile, ultrafast quantum optics has seen remarkable advances in the development of quantum light sources with macroscopic photon occupations. Particular attention was devoted to the squeezed vacuum light \cite{Davidovich1996, Chekhova2015, Andersen2016}, since it represents a scalable resource in terms of photon numbers or ``brightness''. It was clear that the so-called bright squeezed vacuum (BSV) light pulses would open a realm of strong-field physics driven by quantum light, requiring quantum description of light in light-matter interaction, but on the practical side this has remained elusive until recently.

In parallel, there were consistent attempts to develop theories \cite{Cruz2024} that adopt quantum electrodynamics to elaborate the classical driving field as a multi-mode coherent state. In this fully quantum framework, photoemission from the electronic system reveals its non-classical properties: emitted photons representing frequency comb structures spanning the entire HHG frequency spectrum \cite{Gorlach2020}, entanglement between all harmonic field modes \cite{StammerPRL2024, Sloan2023, Lange2024}, and the generation of highly non-classical states such as Schr{\"o}dinger's cat state \cite{Lewenstein2021}. These findings indicated that already classical driving fields can induce quantum photonic effects, and even more intriguing outcomes would emerge when the matter is driven by genuinely quantum light.

Recently, generation of BSV with more than $10^{13}$ photons per few-femtosecond pulse was demonstrated \cite{Chekhova2017, Chekhova2019}. This motivated theoretical proposals to investigate the electron dynamics \cite{Tzur2023, Tzur2024} and HHG driven by non-classical light such as BSV \cite{Gorlach2023, TzurPRR2024}, squeezed coherent light \cite{TzurPRR2024}, or by Fock states \cite{Gorlach2023, Stammer2024}. Then, in 2024, HHG driven by BSV could be studied also experimentally \cite{Chekhova2024}. These developments call now for a deeper theoretical understanding of interacting quantum light and matter with resulting phenomena.

Especially, it is of value to get a more profound and intuitive picture of the tunneling induced by quantum light than currently available following the above-mentioned works. Moreover, it happens that it is possible to avoid approximations in the description of quantum light used in these works, equivalent to assuming of a positive $P$-distribution and neglecting quantum fluctuations with respect to the amplitudes of the involved coherent states \cite{Schleich2001}, paving the way for a more complete quantum picture. In this work, we outline such a formalism, based on an exact decomposition of quantum light into a bundle of deterministic field trajectories.

\begin{figure*}[t]
\includegraphics[width=1.5\columnwidth]{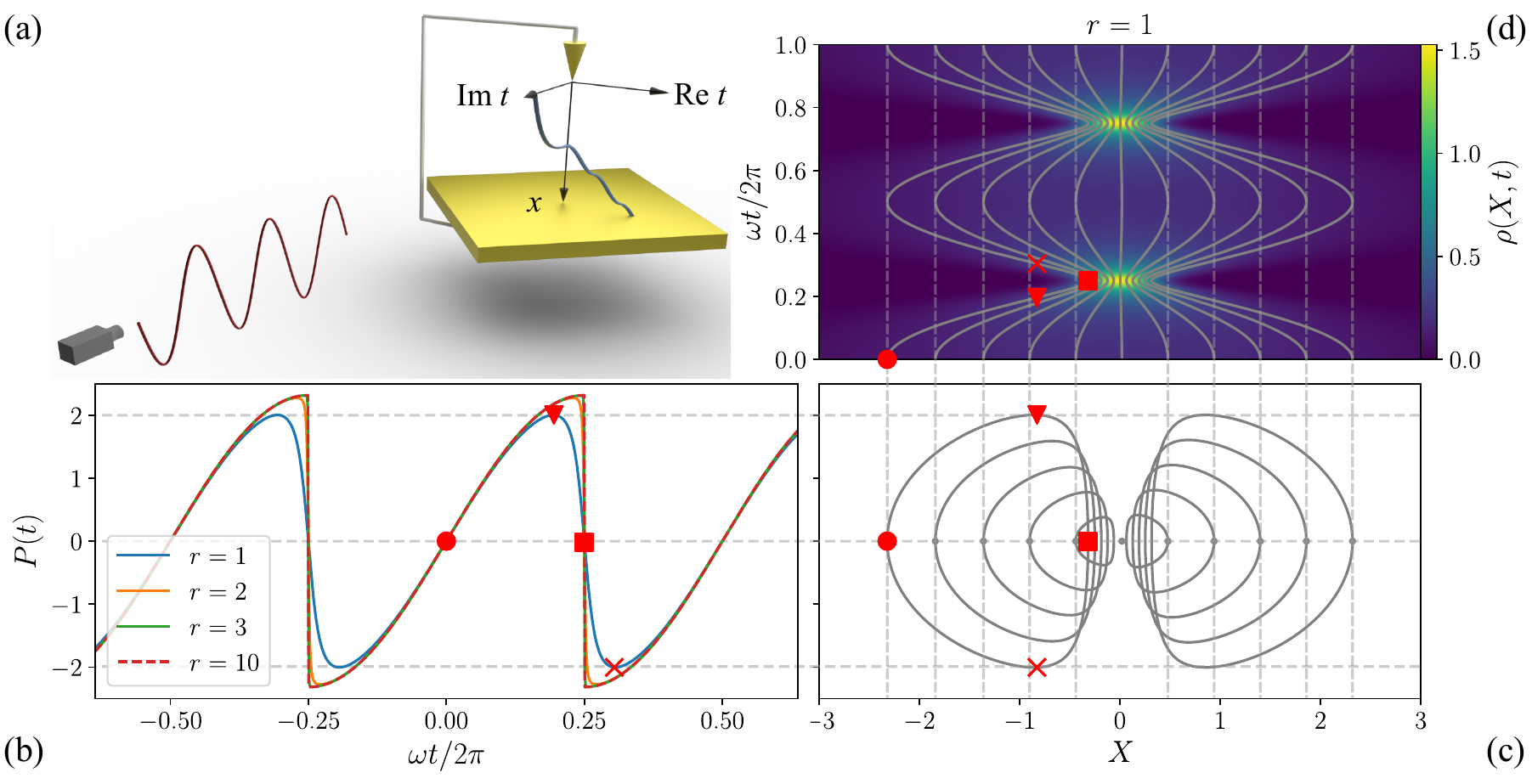}
\caption{(a) Schematic diagram of the setup. A squeezed vacuum state of light drives electronic transport in a metal tip–surface system. The blue curve depicts the complex trajectory of the tunneling electron, starting with an imaginary time under the barrier and going over to a usual classical trajectory in real time as the electron exits into the classically allowed region. (b) Dynamics of the $P$ quadrature (proportional to the electric field) following from Eq.~(\ref{Eq:P_dynamics}) for several values of the squeezing parameter $r$ and an initial realization of the $X$ quadrature, $X(t_i) = -2.32$. (c) Phase-space portrait of the field dynamics for $r=1$. This plot shares the vertical axis with (b) and the horizontal axis with (d). As time evolves, the state trajectory follows a clockwise path, exemplified by the selected realization, where it sequentially passes through the points marked by the circle, triangle, square, and cross. The markers correspond to those in (b) and (d), in the same order. (d) Distribution $\rho(X, t)$ (pseudocolor plot) and Bohmian trajectories $X(t)$ (lines) for the same parameters.
\label{fig1}}
\end{figure*}

\textit{System---}We consider a gold tip and surface with tunneling energy barrier $\Delta U = 5 \, \mathrm{eV}$. The BSV is used to drive the tunneling of electrons through the gap between the tip and surface. See Fig.~\ref{fig1}(a) for the schematic diagram of the setup. This choice of geometry \cite{Kruger2024, Goulielmakis2023, Hommelhoff2024, Siday2024} is to confine the interested direction of tunneling, only collecting one-sided transport of electrons from the tip to the surface. The size of the gap is considered to be a few nanometers.

\textit{Bohmian description of quantum light---}Generally, quantum light can be described either in the Heisenberg picture via its operator properties or in the Schr\"{o}dinger picture representing its state in a certain basis, resulting in the Fock-state and various phase-space representations \cite{Schleich2001, Gorlach2023, Yang2023, Yang2025}. However, with the coupling to quantum matter, operating in the total light-matter Hilbert space is challenging so that approximations were required, as we mentioned above. An elegant way to avoid those approximations, while neglecting backaction to the strong driving field, is provided by the Bohmian formalism \cite{Bohm1,*Bohm2}, which is here, perhaps somewhat surprisingly, applied to capture the quantum dynamics of the light modes as we outline below. Interestingly, this formalism was shown to be useful for investigating strong-field electron dynamics \cite{Jooya2015, Jooya2016}, observing transition between ionization regimes during the HHG process \cite{Li2016} and testing the definitions of tunneling time and exit position \cite{Ivanov2017, Douguet2018, Moon2024, Sharoglazova2025}. However, those works dealt with classical light, whereas it was the electron dynamics that was described by Bohmian trajectories.

For simplicity, we restrict our consideration to a single-frequency mode, corresponding to the central frequency $\omega$ of the applied quantum light pulse. We also assume a fixed linear polarization of the mode. Its Hamiltonian corresponds to a simple harmonic oscillator and reads $\hat{H} = \hbar \omega (\hat{a}^{\dagger} \hat{a} + \frac{1}{2}) = {\hbar \omega}(\hat{P}^2 + \hat{X}^2)/2$, where $\hat{a} = (\hat{X} + i\hat{P}) / \sqrt{2}$ is the annihilation operator. $\hat{X}$ and $\hat{P}$ are operators for two orthogonal quadratures such that $[\hat{X},\hat{P}] = i$. Squeezed vaccum states are introduced by the application of the squeezing operator $\hat{S}(z) = \exp \left[\frac{1}{2} z \hat{a}^{\dagger 2} - \frac{1}{2} z^* \hat{a}^2\right]$ to the vacuum state $|0\rangle$, $|z;0\rangle\equiv\hat{S}(z)|0\rangle$, where $z = r e^{i\phi}$ denotes the squeezing factor and $r>0$. In the $X$ representation for the resulting (stationary) wave function of the squeezed vacuum state in the rotating reference frame (Heisenberg picture), we get
\begin{equation}
    \psi (X) = \psi(0) \exp \left[ - \frac{c_-}{c_+} \frac{1}{2} X^2\right],
\end{equation}
where $c_{\pm} = \cosh r \pm e^{i \phi} \sinh r$. In the Schr\"{o}dinger picture we have $|\psi(t)\rangle = |ze^{-2i\omega t};0\rangle$. The average $\langle \hat{E} \rangle$ of the electric field $\hat{E} = \sqrt{\hbar \omega / \varepsilon_0 V} \hat{P}$ in this state vanishes for all times, whereas its variance $\langle (\hat{E} - \langle \hat{E} \rangle)^{2} \rangle(t) \propto \langle \hat{P}^{2}\rangle(t)$ is finite and oscillates with time. Here, $\varepsilon_{0}$ denotes the vacuum permittivity and $V$ is the quantization volume.

Bohmian mechanics \cite{Bohm1,*Bohm2}, which is also known as a hydrodynamic formulation of quantum mechanics \cite{Madelung1927}, describes the quantum randomness of an observable in terms of a probability distribution. As the state of the light mode propagates, the probability is redistributed along the associated probability flux. In the $X$ representation, the probability distribution is given as $\rho(X,t) = |\psi(X,t)|^{2}$ for the wavefunction $\psi(X,t) = \langle X|\psi(t)\rangle$. The distribution satisfies the continuity equation $\partial_{t}\rho(X,t) = -\partial_{X}J(X,t)$, where $J(X,t) = \omega \mathrm{Im} \left[\psi^* (X,t) \partial_{X} \psi (X,t)\right]$ is the probability flux. As in the usual hydrodynamic theory, the probability flux is proportional to the density and the ``velocity'' of the probability, i.e., $J(X,t)\equiv\rho(X,t)v(X,t)$, so that
\begin{equation}\label{Eq:Xdot}
    \dot{X}(t) = v(X,t) \equiv \frac{J(X,t)}{\rho(X,t)} = \omega \mathrm{Im} \left[\frac{\partial_{X}\psi(X,t)}{\psi(X,t)} \right].
\end{equation}
For a given initial position $X(t_{i})$, which is a random variable following the initial probability distribution $\rho(X,t_{i})$, the trajectory $X(t)$ satisfying Eq. (\ref{Eq:Xdot}) indicates a probability flow line. Redistribution of initial positions along these lines determines the probability distribution $\rho(X,t)$ at any time $t$. The dynamics of the $P$ quadrature, determining the electric field, is found as $P(t) = \dot{X}(t)/\omega$.
For a squeezed vacuum, it is given by \cite{garciachung2025}
\begin{eqnarray}
    P(t) &=& - c(t) X(t_i) e^{- \omega \int_{t_i}^{t} c(t') \, \mathrm{d}t'}, \label{Eq:P_dynamics}\\
    c(t) &=& - \frac{\sin (\phi - 2 \omega t) \sinh (2r)}{\cos (\phi - 2 \omega t) \sinh (2r) + \cosh (2r)}. \label{Eq:c_dynamics}
\end{eqnarray}
Assuming that there is no backaction of the electron to the field, we can select $t_i$ arbitrarily, fixing the initial distribution $\rho (X,t_i) = |\langle X | z e^{-2i \omega t_i} ; 0 \rangle|^2$. Here,
\begin{equation}\label{Eq:rho_Gaussian}
    \rho (X,t) = \left(\frac{c_r (t)}{\pi}\right)^{1/2} e^{- c_r (t) X^2}\;,
\end{equation}
where $c_r(t) = \mathrm{Re} \left[c_- (t)/c_+ (t)\right]$ is given by
\begin{equation}
    c_r(t) = \frac{1}{\cos (\phi - 2 \omega t) \sinh (2r) + \cosh (2r)}\;.
\end{equation}
Figure \ref{fig1}(c) displays a phase-space portrait corresponding to the dynamics of $P(t)$ and $X(t)$. Even for small squeezing, the portrait drastically deviates from that of a classical harmonic oscillator. In particular, $X$ does not change sign over time for any given realization, but it always has a counterpart with the opposite sign. The latter fact is also valid for $P$, even though it exhibits anharmonic oscillations around zero, with the anharmonicity increasing with $r$. Thus, in fact, mean values of both $X$ and $P$ vanish at any time $t$, as expected for purely quantum light.

Based on the parameters relevant to the experimental observation of vacuum fluctuations \cite{Riek2015}, we choose $\omega = 0.0285\, \mathrm{a.u.}$ corresponding to $0.775 \, \mathrm{eV}$, whereas we select $\sqrt{\hbar \omega / \varepsilon_0 V} = \sqrt{2} \times 10^{-8}\,\mathrm{a.u.}$, similar to Ref.~\cite{Tzur2024}. This unravels the squeezed vacuum state into a bundle of classical realizations of the electric field $E(t)=\sqrt{\hbar \omega / \varepsilon_0 V}P(t)$, such as illustrated in Fig.~\ref{fig1}(b). As $r$ increases, their temporal shape becomes sawtooth with abrupt changes around time moments $t_n$ satisfying $\omega t_n / 2 \pi = \left(2n+1\right)/4$, $n \in \mathbb{Z}$.

\textit{Quasiclassical non-adiabatic tunneling theory---}When the BSV field is applied to induce electronic tunneling, we can consider the corresponding process as resulting from tunneling events caused by each of the classical realizations of the field from the bundle. It is known that tunneling caused by classical fields, e.g., within the three-step model, can be typically well captured by a quasiclassical description. In the non-adiabatic regime, when the electron gains energy as moving under the barrier, this can be described by a generalization of the WKB theory \cite{LandauQM} provided by the imaginary time method \cite{Perelomov1966,Popov2005}. However, this method did not work for the asymmetric field profiles as in Fig.~\ref{fig1}(b), in contrast to a more general formulation developed in Ref.~\cite{Kim2021} that we apply below.

The quasiclassical wavefunction of an electron at a position $x$, and a time $t$ is approximated by
\begin{equation}\label{Eq:ansatz}
    \psi_\mathrm{el} (x,t) \propto e^{\frac{i}{\hbar} S(x,t)},
\end{equation}
where
the action
\begin{equation}
    S (x,t) = \int_{t_0}^{t} \mathcal{L}(x', \dot{x}', t') dt' - \mathcal{E} t_0
\end{equation}
represents the general solution \cite{LandauQM, Epstein1964} of the Hamilton-Jacobi equation. Whereas $x$ and $t$ are real, $t_0$ is generally complex, as well as $x'$ and $t'$. Here $\mathcal{E} = \mathcal{H}|_{t=t_0}$ is the Hamiltonian of the electron at time $t_0$, when the electron is starting its motion under the energy barrier inside the gap. Since the kinetic energy there is negative, the electron velocity is necessarily complex. We assume a constant static potential barrier $U(x') = -\Delta U \Theta(-x')$, where $\Theta(x)$ denotes the Heaviside step function, in the relevant spatial region above the surface and a classical realization of the quantum light $E(t)$ enters the Lagrangian $\mathcal{L}$ as a driving field so that
\begin{equation}\label{Eq:Lagrangian}
    \mathcal{L}(x', \dot{x}', t') = \frac{1}{2} m \dot{x}'^2 + e x' E(t') - U(x'),
\end{equation}
where $e$ denotes the electron charge and $m$ is the mass of the electron.

\begin{figure}[t]
\includegraphics[width=\columnwidth]{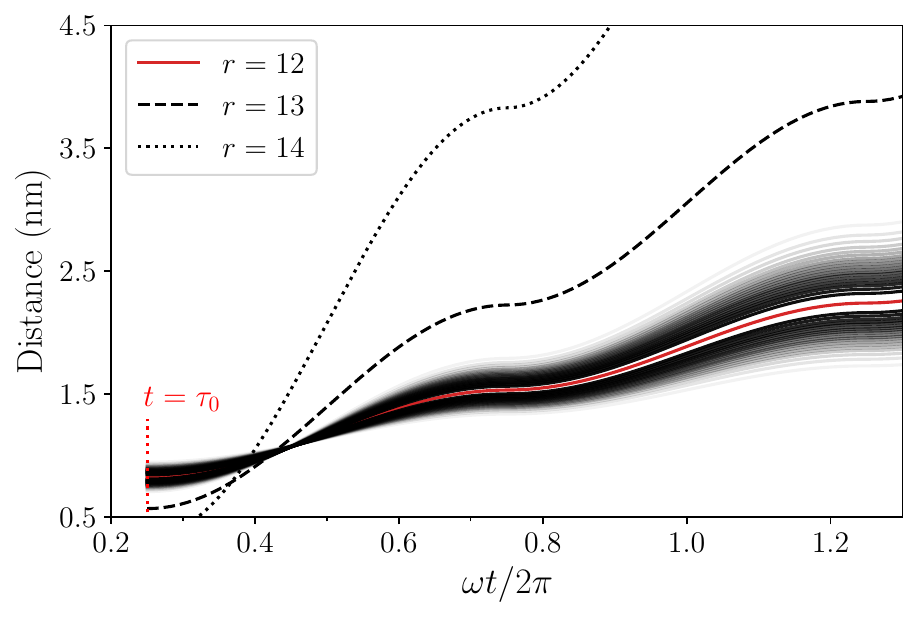}
\caption{Trajectories of the released electrons in the classically allowed region for several values of the squeezing factor $r$ and the optimal initial realization $X(t_i) = X_\mathrm{peak}$ (visualized in Fig.~\ref{fig2}). Full red line shows the trajectory for $r=12$ and $X(t_i) = X_\mathrm{peak}$, whereas the set of trajectories corresponds to less optimal realizations $X(t_i)=X_l$ defined by $\rho(X_l) \mathcal{P}(X_l) = (1-l/20) \rho(X_{\mathrm{peak}}) \mathcal{P}(X_{\mathrm{peak}})$ for $l = 1,2, \cdots, 19$. As $l$ increases (reflected in the color fading of the lines), the deviations from the trajectory with $X(t_i) = X_\mathrm{peak}$ grow. Red dotted line indicates $t=\tau_0$ for all of the trajectories.
\label{fig4}}
\end{figure}

Euler-Lagrange equations of motion (EOMs) determined by $\mathcal{L}$ need to be supplemented by additional conditions to get a closed solution, keeping in mind that $t_0$ is also unknown and that the relevant quantities are complex-valued. Two of them (four in terms of real variables) are boundary conditions:
\begin{equation}\label{Eq:BC}
    x'(t'=t_0) = 0, \quad x'(t'=t) = x.
\end{equation}
Since at $t=t_0$ under the barrier we have $\mathcal{H}|_{t = t_0} = - \Delta U$, additionally we have condition $\dot{x}' (t=t_0) = i \sqrt{2 \Delta U / m}$ (implying two conditions in terms of real variables). Solving the EOMs with the above conditions allows finding of $S(x,t)$ as well as $t_0(x,t)$. However, when we are interested in the tunneling probability with exponential accuracy, we should additionally demand $\partial_t \mathrm{Im} \, S(x,t)=0$ at any relevant $x$ and $t$, meaning $\mathrm{Im} [m\dot{x}'(t'=t)]=0$ in the classically allowed region after the barrier \cite{Moskalenko1999,Moskalenko2000,Kim2021}. Since $\mathrm{Im}\,x=0$ and we may also select $t=0$ due to $\partial_t \mathrm{Im} \, S(x,t)=0$, following from the Hamilton-Jacobi equation, this extra condition fixes the values of $t_0$ and $S\equiv S_\mathrm{opt}$. As a result, with exponential accuracy, we can determine the tunneling probability as $\mathcal{P} = e^{-\frac{2}{\hbar} \mathrm{Im} \, S_\mathrm{opt}}$.

\begin{figure}[t]
\includegraphics[width=\columnwidth]{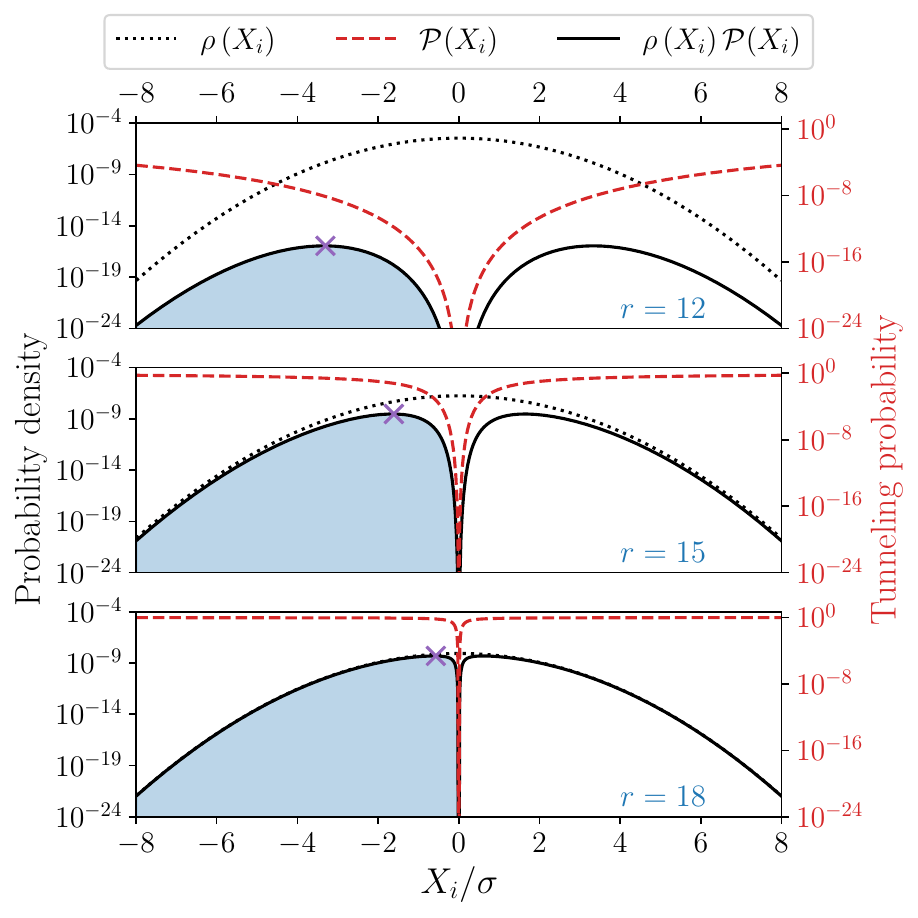}
\caption{Probability distribution $\rho \left(X_i\right)$ of initial realizations for the $X$ quadrature, tunneling probability $\mathcal{P} \left(X_i\right)$ and $\rho \left(X_i\right) \mathcal{P} \left(X_i\right)$ for several values of the squeezing factor. $\sigma$ denotes the standard deviation of $\rho(X_i)$, points $X_i=X_\mathrm{peak}$ where $\rho \left(X_i\right) \mathcal{P} \left(X_i\right)$ is maximized are indicated by purple crosses and the area corresponding to the integral in Eq.~(\ref{Eq:P_tot}) is colored in blue.
\label{fig2}}
\end{figure}

We find that the real time moments $\tau_0 = \mathrm{Re} \, t_0$, at which we can think the real optimal trajectory to emerge after the barrier, are very close to the vertical edges of the temporal profile of the electric field [cf. Fig.~\ref{fig1}(b)]:
\begin{equation}
    \tau_0=\frac{2\pi}{\omega}\left[\frac{1}{4} +\varepsilon+\frac{n}{2}\right], \quad n \in \mathbb{Z}\;.
\end{equation}
Here $0< \varepsilon \ll 1$ for all relevant realizations and considered values of $r$.
Due to the periodicity, it is sufficient to limit the consideration to $n=0$. $\tau_0$ does not coincide with the position of the peak of the field, being shifted toward the nodal point with respect to it. We took into account that because of the asymmetric geometrical configuration of the system, only the electrons propagating from the tip to the surface have to be considered, but not vice versa. This means that the force experienced by the electron $F=- |e| E(\tau_0)$ at $t=\tau_0$ should be positive, implying $E(\tau_{0})\propto P(\tau_0)<0$. Additionally, one has to take into account that there are two types of realizations depending on the sign of $X$ [cf. Fig.~\ref{fig1}(c)]: $X>0$ and $X<0$. Only the case $X<0$, as selected in Fig.~\ref{fig1}(b), matters because otherwise the force flips its sign shortly after the emergence of the electron and it is immediately dragged back towards the tip. In contrast, for $X<0$ the temporal profile is such that the electron in the classical region propagates toward the surface without changing the direction of motion, as can be seen in Fig.~\ref{fig4}.

\begin{figure}[t]
\includegraphics[width=\columnwidth]{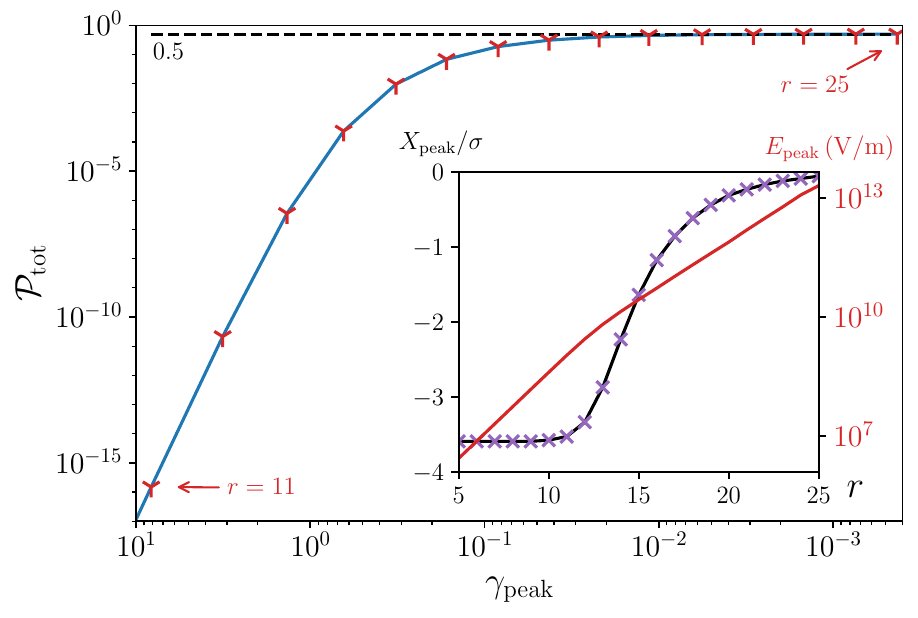}
\caption{Dependence of the total tunneling probability of the electron $\mathcal{P}_\mathrm{tot}$ evaluated based on Eq.~(\ref{Eq:P_tot}) on the effective Keldysh parameter $\gamma_{\mathrm{peak}}$ introduced in Eq.~(\ref{Eq:gamma}). From left to right, the corresponding squeezing factor increases from $r = 11$ to $r = 25$. The inset illustrates the dependences of $X_{\mathrm{peak}}$ and $E_{\mathrm{peak}}$ on $r$, where purple markers are shared with Fig.~\ref{fig2}.
\label{fig3}}
\end{figure}

\textit{Trajectory-averaged tunneling probability---}To find the
resulting probability of tunneling $\mathcal{P}_\mathrm{tot}$ induced by the quantum light,  we need to average the tunneling probability for each particular field realization $\mathcal{P}(X_i)$ over the probability distribution $\rho(X_i)$:
\begin{equation}\label{Eq:P_tot}
    \mathcal{P}_\mathrm{tot}=\int_{-\infty}^0  \mathcal{P}(X_i) \rho(X_i) \, \mathrm{d}X_i,
\end{equation}
where we took only negative $X_i$ into account, as explained above. According to Eq.~(\ref{Eq:rho_Gaussian}), $\rho(X_i)$ is Gaussian. We denote its standard deviation as $\sigma=e^{-r}$. Dependence of $\mathcal{P}(X_i)$, $\rho(X_i)$ and their product on $X_i$ are shown in Fig.~\ref{fig2} for the interval within $8\sigma$. Whereas $\mathcal{P}(X_i)$ increases with the magnitude of $X_i$, $\rho(X_i)$ decreases, so that the function $\mathcal{P}(X_i) \rho(X_i)$ integrated in Eq.~(\ref{Eq:P_tot}) has a maximum at a certain value of $X_i$, which we denote as $X_\mathrm{peak}$. $E_\mathrm{peak}$ is the maximum value of $E$ corresponding to this realization, reached periodically with period $\pi/\omega$. As $r$ increases, $X_{\mathrm{peak}} / \sigma$ becomes smaller but never reaches $0$. In contrast, $E_{\mathrm{peak}}$ increases exponentially with $r$, as can be expected (see inset of Fig.~\ref{fig3}). In connection with that, it is useful to introduce the effective Keldysh parameter \cite{Keldysh1965,Popov2004review,Popruzhenko2014} as
\begin{equation}\label{Eq:gamma}
    \gamma_{\mathrm{peak}} = \frac{\omega \sqrt{2 m \Delta U}}{e E_{\mathrm{peak}}}\;.
\end{equation}

Dependence of $\mathcal{P}_\mathrm{tot}$ on $\gamma_{\mathrm{peak}}$ and $r$, calculated after Eq.~(\ref{Eq:P_tot}), is depicted in Fig.~\ref{fig3}. We can recall that for classical harmonic driving, values of the Keldysh parameter $\gamma$ allow to classify the induced process as direct tunneling ($\gamma\ll1$), multiphoton transition ($\gamma\gg 1$) or non-adiabatic, photon-assisted tunneling bridging both regimes. We can also see in Fig.~\ref{fig3} that as $r$ and $E_{\mathrm{peak}}$ increase, whereas $\gamma_{\mathrm{peak}}$ correspondingly decreases and other parameters are fixed, the dependence $\mathcal{P}_\mathrm{tot}(\gamma_{\mathrm{peak}})$ smoothly changes its character when $\gamma_{\mathrm{peak}}$ is on the order of unity. Thus, we see that the classification based on the value of the Keldysh parameter can be generalized to the drivings by quantum light.

\textit{Conclusion and outlook---}We have demonstrated that representing quantum light as a bundle of deterministic Bohmian field trajectories enables a rigorous framework for treating such light in light-matter interaction problems. In particular, our results show that this approach can be excellently combined with the non-adiabatic quasiclassical description of tunneling, generalizing the corresponding description established for the case of classical light. Especially in cases of asymmetry in terms of the tunneling direction, as for the tunneling-microscope or asymmetric nano-antenna configurations, our theory provides an efficient solution path with a clear view of the process. It may serve as a platform to consider optically-induced tunneling experiments driven with arbitrary quantum states of light. Moreover, our framework to treat interaction with quantum light is capable of bridging any semi-classical theories, such as the time-dependent Schr\"odinger equation or its Bohmian counterpart, which describe electron motion under classical driving, with their fully quantum analogues. Further development of the presented approach should open up the possibility of studying the interaction between quantum light and matter from various perspectives, such as driving with multi-mode quantum light or generating quantum entanglement between light and matter.

\begin{acknowledgments}
This research was supported by the KAIST C2 and UP Projects. D.V.S. acknowledges funding by the European Union's Horizon Europe Research and Innovation Programme under agreement 101070700 (project MIRAQLS). A.S.M. acknowledges funding by the Deutsche Forschungsgemeinschaft (DFG)--Project No. 425217212--SFB 1432.
\end{acknowledgments}

\bibliography{bohmquasi}

\end{document}